# Architecture of Network and Client-Server model

Hai Zhang

University of Science and Technology of China

**Abstract** –With the development of Internet technology, the Web is becoming more and more important in our lives so that it has even become an essential element. The application of the Web has never been limited to computers; it has been opened to all kinds of intelligent digital devices like mobile ones. In this paper, we discuss the main and substantial difference between UDP and TCP, and how to implement Client-Server model. We also discuss the efficiency of multi-thread server and its relationship with internet.

## 1. Introduction

With the development of Internet technology, the Web is becoming more and more important in our lives so that it has even become an essential element. At the same time, the application of the Web has never been limited to computers; it has been opened to all kinds of intelligent digital devices like mobile ones. In addition, the architecture of the Web is the Client-Sever (CS) model that is shown as Figure 1. As a result, the communication between server and client is the first thing we should be concerned about. Only based on successful and smooth communication, can Web technology move forward and Web applications be applied to all kinds of devices so that they can help people.

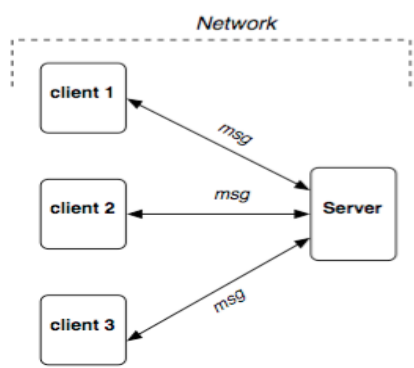

**Figure 1** The structure of the Client-Server model.

## 2. The comparison between TCP and UDP

To implement communication, firstly, we should define how to transfer messages between server and client, that is, the protocol that is one kind of language that can be understood by terminals, which connect to Internet. Just as we learned one new language including its syntax and sentence structure, so that we can make a complete sentence and paragraph, for terminal communication, this language is called protocol. The first part that I want to introduce concerns different transfer protocols used between client and server for different purposes. Secondly, as long as a terminal knows the language used by communication, when it has something to transfer, it can create a packet based on its language and protocol. Of course, in the concrete implementation of communication, especially, when we want to achieve it by programming, packet is not enough for transmission, we need another structure to finish packing our message that is called a socket. The socket is the basic operating unit that sustains network communication.

In this semester, one of the courses I am taking is Computer network which introduces the transfer protocol between different terminals. Just as I mentioned above briefly, a protocol can be defined as the rules, conventions, and standards governing the syntax, semantics, and synchronization of communication. In other words, protocols are sets of rules, (or a sequence of events) that control or enable preferably reliable and recognizable transfer of information among communication end points**.** There are so many transfer protocols used in communication because of different purposes. That is to say, devices may be connected but not communicate without protocol. Here, I will introduce the two most common protocols that are "User Datagram Protocol (UDP) and Transmission Control Protocol (TCP)". UDP does not have prior communication to set up a special transmission channel so that it is a simple transmission model with a minimum of protocol mechanism. In addition, this is the main and substantial difference between UDP and TCP. For TCP, before two terminals establish a connection, they will execute a process called handshaking that is a process of negotiation. After handshaking, the two terminals will reach a agreement including, but not limited to, information transfer rate, coding alphabet, parity, interrupt procedure, and other protocol or hardware features. As a result of handshaking, TCP will establish a specific connection between two terminals for this communication, so that TCP will cost more time depends on network traffic condition than UDP to finish it, however, TCP is simultaneously much more reliable than UDP. So whether our transmission uses TCP or UDP is a trade off, we have to consider network overload and the degree of security.

| TCP | UDP |
|---|---|
| Reliable | Unreliable |
| Connection-oriented | Connectionless |
| Segment retransmission and flow control through windowing | No windowing or retransmission |
| Segment sequencing | No sequencing |
| Acknowledge segments | No acknowledgement |

**Figure 2** The comparison between TCP and UDP

## 3. How to use Client-Server model into real world

After the above introduction of transfer protocol, we have to begin to be concerned with how to put these protocols into real use, that is to say, how to implement network communication by programming. The most common method is by socket, which is network interface, an endpoint of an inter-process communication flow across a computer network. Actually, there are many kinds of sockets, but nowadays, most communications between computers are based on the Internet protocol that this paper introduces so that most network sockets are called Internet sockets. A socket contains socket address and data. Socket address includes local IP address and remote IP address that are the addresses of our computer and target computer in the network and port number that is used for sending and receiving data. About data, it is not only something the computer wants to transfer, but also some elements used for synchronization that is a part of the protocol. However, we can not operate sockets directly because it is just abstract concept, therefore, we have to through application programming interface (API) to operate sockets including sending, receiving data and functions. It is worth noting that the API is not identical among different platforms and programming languages. I am working on one computer communication project by programming C and the other android project that also includes communication between one server and many clients using Java. Even though all of Internet protocols I used are TCP, the achievement is different with each other, programming C needs more complicated statements because it is more basic than Java, Java API encapsulates more operations, but I have to implement many functions by myself in programming C, which is also the difference between oriented-object language and structure programming language.

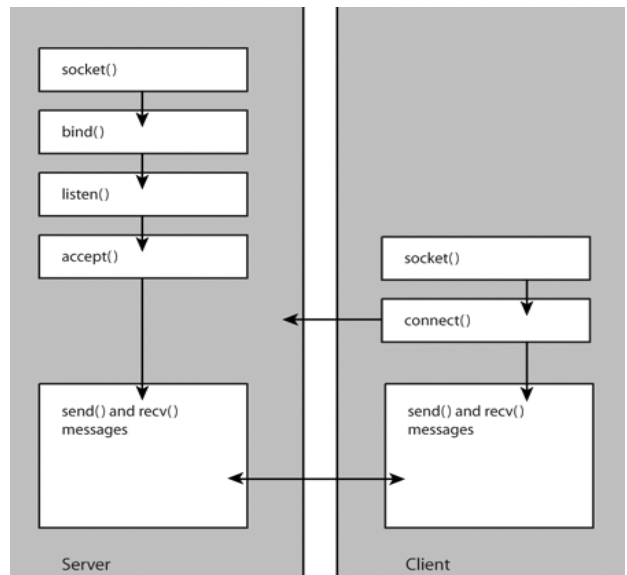

**Figure 3** TCP Socket Connections

## 4. Discussion

In addition, if I have finished the communication between one server and several clients, the other factor we need to consider is efficiency. Because I need to deal with several clients, if we handle requests one by one, the total cost time will be N times by one execution time. However, if we can do it parallel, the total time can be almost equal to one execution time. Another course I am taking this semester is Operating System, which introduces how to deal with some different questions or different parts in one question by multi-thread. As a result, when I designed a pattern of one server that has to decrease the overload due to too many clients, I decided to achieve it by multi-thread on server side. To be specific, when one client wants to connect to the server, the server will open a new thread for it, this thread just belongs to the client and it will handle every request. Meanwhile, in my project, there is a function like MSN so that it needs to communicate between two different threads. To do it, the server will store its specific ID and its thread for every client so that the server can find the target thread by its ID and then send message to it from the original client. In addition, what I want to say is multi-thread can be used for solving different parts in one question. The most basic one is to calculate the product of two matrixes. Of course, if the dimension of matrix is small, we do not need to worry about it, we just process it row-by-row and column-by-column. However, if they are too large, this method will cost much time, therefore, we can assign one thread to one row and one column, and then we can improve efficiency by calculating every row and column simultaneously.

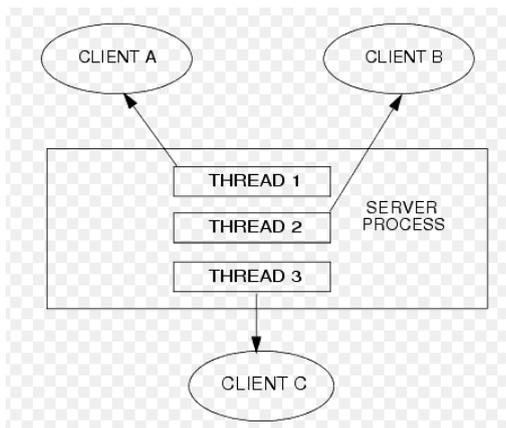

Figure 4 Multi-thread processes different clients

In a word, if we want to design and finish a network project practically, the first thing is to understand Internet protocol and secondly, we pack it by socket and send it. Finally, we should consider the efficiency of the server so that multi-thread would be imported into project.